\documentclass[reprint,amsmath,amssymb,aps,prl,superscriptaddress]{revtex4-2}

\usepackage{graphicx}
\usepackage{amsmath}
\usepackage{dcolumn}
\usepackage{bm}
\usepackage{xcolor}
\usepackage{titlesec}
\usepackage{float}
\usepackage{tabularx}
\usepackage[T1]{fontenc}
\usepackage[colorlinks=true, citecolor=LinkBlue, linkcolor=LinkBlue, urlcolor = LinkBlue]{hyperref}
\usepackage{lineno}

\makeatletter
\g@addto@macro\bfseries{\boldmath}
\makeatother
\definecolor{LinkBlue}{HTML}{2f3093}
\titleformat{\section}[runin]{\it }{}{0pt}{\indent}[.---\kern-1.8em]
\titlespacing*{\section}{0pt}{1.5ex}{4.5ex}
\makeatletter
\def\@fnsymbol#1{\ensuremath{\ifcase#1\or
   *\or \dagger\or \ddagger\or \mathsection\or \mathparagraph\or \|\or
   **\or \dagger\dagger\or \ddagger\ddagger\or
   \mathsection\mathsection\or \mathparagraph\mathparagraph\or \|\|\or
   ***\or \dagger\dagger\dagger\or \ddagger\ddagger\ddagger\or
   \mathsection\mathsection\mathsection\or \mathparagraph\mathparagraph\mathparagraph\or \|\|\|\or
   ****\or \dagger\dagger\dagger\dagger\or \ddagger\ddagger\ddagger\ddagger
   \else\@ctrerr\fi}}
\makeatother

\begin{document}

\preprint{APS/123-QED}

\title{First Measurement of the Electron Neutrino Charged-Current Pion Production Cross Section on Carbon with the T2K Near Detector\\}

\newcommand{\INSTHD}{\affiliation{University Autonoma Madrid, Department of Theoretical Physics, 28049 Madrid, Spain}}
\newcommand{\INSTFE}{\affiliation{Boston University, Department of Physics, Boston, Massachusetts, U.S.A.}}
\newcommand{\INSTD}{\affiliation{University of British Columbia, Department of Physics and Astronomy, Vancouver, British Columbia, Canada}}
\newcommand{\INSTGA}{\affiliation{University of California, Irvine, Department of Physics and Astronomy, Irvine, California, U.S.A.}}
\newcommand{\INSTI}{\affiliation{IRFU, CEA, Universit\'e Paris-Saclay, F-91191 Gif-sur-Yvette, France}}
\newcommand{\INSTGB}{\affiliation{University of Colorado at Boulder, Department of Physics, Boulder, Colorado, U.S.A.}}
\newcommand{\INSTFH}{\affiliation{Duke University, Department of Physics, Durham, North Carolina, U.S.A.}}
\newcommand{\INSTJA}{\affiliation{E\"{o}tv\"{o}s Lor\'{a}nd University, Department of Atomic Physics, Budapest, Hungary}}
\newcommand{\INSTEF}{\affiliation{ETH Zurich, Institute for Particle Physics and Astrophysics, Zurich, Switzerland}}
\newcommand{\INSTIG}{\affiliation{VNU University of Science, Vietnam National University, Hanoi, Vietnam}}
\newcommand{\INSTIE}{\affiliation{CERN European Organization for Nuclear Research, CH-1211 Gen\'eve 23, Switzerland}}
\newcommand{\INSTEG}{\affiliation{University of Geneva, Section de Physique, DPNC, Geneva, Switzerland}}
\newcommand{\INSTHJ}{\affiliation{University of Glasgow, School of Physics and Astronomy, Glasgow, United Kingdom}}
\newcommand{\INSTJG}{\affiliation{Ghent University, Department of Physics and Astronomy, Proeftuinstraat 86, B-9000 Gent, Belgium}}
\newcommand{\INSTDG}{\affiliation{H. Niewodniczanski Institute of Nuclear Physics PAN, Cracow, Poland}}
\newcommand{\INSTCB}{\affiliation{High Energy Accelerator Research Organization (KEK), Tsukuba, Ibaraki, Japan}}
\newcommand{\INSTIB}{\affiliation{University of Houston, Department of Physics, Houston, Texas, U.S.A.}}
\newcommand{\INSTED}{\affiliation{Institut de Fisica d'Altes Energies (IFAE) - The Barcelona Institute of Science and Technology, Campus UAB, Bellaterra (Barcelona) Spain}}
\newcommand{\INSTJC}{\affiliation{Institut f\"ur Physik, Johannes Gutenberg-Universit\"at Mainz, Staudingerweg 7, 55128 Mainz, Germany}}
\newcommand{\INSTHH}{\affiliation{Institute For Interdisciplinary Research in Science and Education (IFIRSE), ICISE, Quy Nhon, Vietnam}}
\newcommand{\INSTEI}{\affiliation{Imperial College London, Department of Physics, London, United Kingdom}}
\newcommand{\INSTGF}{\affiliation{INFN Sezione di Bari and Universit\`a e Politecnico di Bari, Dipartimento Interuniversitario di Fisica, Bari, Italy}}
\newcommand{\INSTBE}{\affiliation{INFN Sezione di Napoli and Universit\`a di Napoli, Dipartimento di Fisica, Napoli, Italy}}
\newcommand{\INSTBF}{\affiliation{INFN Sezione di Padova and Universit\`a di Padova, Dipartimento di Fisica, Padova, Italy}}
\newcommand{\INSTBD}{\affiliation{INFN Sezione di Roma and Universit\`a di Roma ``La Sapienza'', Roma, Italy}}
\newcommand{\INSTEB}{\affiliation{Institute for Nuclear Research of the Russian Academy of Sciences, Moscow, Russia}}
\newcommand{\INSTHI}{\affiliation{International Centre of Physics, Institute of Physics (IOP), Vietnam Academy of Science and Technology (VAST), 10 Dao Tan, Ba Dinh, Hanoi, Vietnam}}
\newcommand{\INSTJD}{\affiliation{ILANCE, CNRS – University of Tokyo International Research Laboratory, Kashiwa, Chiba 277-8582, Japan}}
\newcommand{\INSTHA}{\affiliation{Kavli Institute for the Physics and Mathematics of the Universe (WPI), The University of Tokyo Institutes for Advanced Study, University of Tokyo, Kashiwa, Chiba, Japan}}
\newcommand{\INSTID}{\affiliation{Keio University, Department of Physics, Kanagawa, Japan}}
\newcommand{\INSTIF}{\affiliation{King's College London, Department of Physics, Strand, London WC2R 2LS, United Kingdom}}
\newcommand{\INSTCC}{\affiliation{Kobe University, Kobe, Japan}}
\newcommand{\INSTCD}{\affiliation{Kyoto University, Department of Physics, Kyoto, Japan}}
\newcommand{\INSTEJ}{\affiliation{Lancaster University, Physics Department, Lancaster, United Kingdom}}
\newcommand{\INSTII}{\affiliation{Lawrence Berkeley National Laboratory, Berkeley, California, U.S.A.}}
\newcommand{\INSTBA}{\affiliation{Ecole Polytechnique, IN2P3-CNRS, Laboratoire Leprince-Ringuet, Palaiseau, France}}
\newcommand{\INSTFC}{\affiliation{University of Liverpool, Department of Physics, Liverpool, United Kingdom}}
\newcommand{\INSTFI}{\affiliation{Louisiana State University, Department of Physics and Astronomy, Baton Rouge, Louisiana, U.S.A.}}
\newcommand{\INSTIH}{\affiliation{Joint Institute for Nuclear Research, Dubna, Moscow Region, Russia}}
\newcommand{\INSTHB}{\affiliation{Michigan State University, Department of Physics and Astronomy,  East Lansing, Michigan, U.S.A.}}
\newcommand{\INSTCE}{\affiliation{Miyagi University of Education, Department of Physics, Sendai, Japan}}
\newcommand{\INSTDF}{\affiliation{National Centre for Nuclear Research, Warsaw, Poland}}
\newcommand{\INSTFJ}{\affiliation{State University of New York at Stony Brook, Department of Physics and Astronomy, Stony Brook, New York, U.S.A.}}
\newcommand{\INSTEH}{\affiliation{STFC, Rutherford Appleton Laboratory, Harwell Oxford,  and  Daresbury Laboratory, Warrington, United Kingdom}}
\newcommand{\INSTGJ}{\affiliation{Okayama University, Department of Physics, Okayama, Japan}}
\newcommand{\INSTCF}{\affiliation{Osaka Metropolitan University, Department of Physics, Osaka, Japan}}
\newcommand{\INSTGG}{\affiliation{Oxford University, Department of Physics, Oxford, United Kingdom}}
\newcommand{\INSTIC}{\affiliation{University of Pennsylvania, Department of Physics and Astronomy,  Philadelphia, Pennsylvania, U.S.A.}}
\newcommand{\INSTGC}{\affiliation{University of Pittsburgh, Department of Physics and Astronomy, Pittsburgh, Pennsylvania, U.S.A.}}
\newcommand{\INSTGD}{\affiliation{University of Rochester, Department of Physics and Astronomy, Rochester, New York, U.S.A.}}
\newcommand{\INSTHC}{\affiliation{Royal Holloway University of London, Department of Physics, Egham, Surrey, United Kingdom}}
\newcommand{\INSTBC}{\affiliation{RWTH Aachen University, III. Physikalisches Institut, Aachen, Germany}}
\newcommand{\INSTJF}{\affiliation{School of Physics and Astronomy, University of Minnesota, Minneapolis, Minnesota, U.S.A.}}
\newcommand{\INSTJB}{\affiliation{Departamento de F\'isica At\'omica, Molecular y Nuclear, Universidad de Sevilla, 41080 Sevilla, Spain}}
\newcommand{\INSTFB}{\affiliation{University of Sheffield, School of Mathematical and Physical Sciences, Sheffield, United Kingdom}}
\newcommand{\INSTDI}{\affiliation{University of Silesia, Institute of Physics, Katowice, Poland}}
\newcommand{\INSTIA}{\affiliation{SLAC National Accelerator Laboratory, Stanford University, Menlo Park, California, U.S.A.}}
\newcommand{\INSTBB}{\affiliation{Sorbonne Universit\'e, CNRS/IN2P3, Laboratoire de Physique Nucl\'eaire et de Hautes Energies (LPNHE), Paris, France}}
\newcommand{\INSTJE}{\affiliation{South Dakota School of Mines and Technology, 501 East Saint Joseph Street, Rapid City, SD 57701, United States}}
\newcommand{\INSTCH}{\affiliation{University of Tokyo, Department of Physics, Tokyo, Japan}}
\newcommand{\INSTBJ}{\affiliation{University of Tokyo, Institute for Cosmic Ray Research, Kamioka Observatory, Kamioka, Japan}}
\newcommand{\INSTCG}{\affiliation{University of Tokyo, Institute for Cosmic Ray Research, Research Center for Cosmic Neutrinos, Kashiwa, Japan}}
\newcommand{\INSTHF}{\affiliation{Institute of Science Tokyo, Department of Physics, Tokyo}}
\newcommand{\INSTGI}{\affiliation{Tokyo Metropolitan University, Department of Physics, Tokyo, Japan}}
\newcommand{\INSTHG}{\affiliation{Tokyo University of Science, Faculty of Science and Technology, Department of Physics, Noda, Chiba, Japan}}
\newcommand{\INSTB}{\affiliation{TRIUMF, Vancouver, British Columbia, Canada}}
\newcommand{\INSTJH}{\affiliation{University of Toyama, Department of Physics, Toyama, Japan}}
\newcommand{\INSTDJ}{\affiliation{University of Warsaw, Faculty of Physics, Warsaw, Poland}}
\newcommand{\INSTDH}{\affiliation{Warsaw University of Technology, Institute of Radioelectronics and Multimedia Technology, Warsaw, Poland}}
\newcommand{\INSTIJ}{\affiliation{Tohoku University, Faculty of Science, Department of Physics, Miyagi, Japan}}
\newcommand{\INSTFD}{\affiliation{University of Warwick, Department of Physics, Coventry, United Kingdom}}
\newcommand{\INSTEA}{\affiliation{Wroclaw University, Faculty of Physics and Astronomy, Wroclaw, Poland}}
\newcommand{\INSTHE}{\affiliation{Yokohama National University, Department of Physics, Yokohama, Japan}}
\newcommand{\INSTH}{\affiliation{York University, Department of Physics and Astronomy, Toronto, Ontario, Canada}}

\INSTHD
\INSTFE
\INSTD
\INSTGA
\INSTI
\INSTGB
\INSTFH
\INSTJA
\INSTEF
\INSTIG
\INSTIE
\INSTEG
\INSTHJ
\INSTJG
\INSTDG
\INSTCB
\INSTIB
\INSTED
\INSTJC
\INSTHH
\INSTEI
\INSTGF
\INSTBE
\INSTBF
\INSTBD
\INSTEB
\INSTHI
\INSTJD
\INSTHA
\INSTID
\INSTIF
\INSTCC
\INSTCD
\INSTEJ
\INSTII
\INSTBA
\INSTFC
\INSTFI
\INSTIH
\INSTHB
\INSTCE
\INSTDF
\INSTFJ
\INSTEH
\INSTGJ
\INSTCF
\INSTGG
\INSTIC
\INSTGC
\INSTGD
\INSTHC
\INSTBC
\INSTJF
\INSTJB
\INSTFB
\INSTDI
\INSTIA
\INSTBB
\INSTJE
\INSTCH
\INSTBJ
\INSTCG
\INSTHF
\INSTGI
\INSTHG
\INSTB
\INSTJH
\INSTDJ
\INSTDH
\INSTIJ
\INSTFD
\INSTEA
\INSTHE
\INSTH

\author{K.\,Abe}\INSTBJ
\author{S.\,Abe}\INSTBJ
\author{R.\,Akutsu}\INSTCB
\author{H.\,Alarakia-Charles}\INSTEJ
\author{Y.I.\,Alj Hakim}\INSTFB
\author{S.\,Alonso Monsalve}\INSTEF
\author{L.\,Anthony}\INSTEI
\author{S.\,Aoki}\INSTCC
\author{K.A.\,Apte}\INSTEI
\author{T.\,Arai}\INSTCH
% ...

\collaboration{The T2K Collaboration}\noaffiliation

\begin{abstract}
The T2K Collaboration presents the first measurement of electron neutrino-induced charged-current pion production on carbon in a restricted kinematical phase space. This is performed using data from the $2.5^{\circ}$ off-axis near detector, ND280. The differential cross sections with respect to the outgoing electron and pion kinematics, in addition to the total flux-integrated cross section, are obtained. Comparisons between the measured and predicted cross section results using the \textsc{Neut}, \textsc{Genie} and \textsc{NuWro} Monte Carlo event generators are presented. The measured total flux-integrated cross section is $[2.52 \pm 0.52\;(\text{stat}) \pm 0.30\;(\text{sys})] \times 10^{-39}\;\text{cm}^{2}\;\text{nucleon}^{-1}$, which is lower than the event generator predictions.
\end{abstract}

\maketitle

\section{Introduction}
\label{sec:Introduction}

Charged-current (CC) pion production from electron neutrinos scattering off nucleons $\left( \nu_e \mathrm{CC} \pi^+ \right)$ is a sub-dominant interaction that contributes to the $\nu_e$ appearance signal in accelerator-based long-baseline neutrino oscillation experiments such as T2K and NOvA~\cite{T2K-nue-appearance, Nova-nue-appearance}. This channel is sensitive to the parameters that favour $\nu_\mu \rightarrow \nu_e$ oscillations, including $\delta_{\text{CP}}$ and $\theta_{13}$. The modelling of neutrino-nucleon interactions contributes ${\sim}3\%$ to the total systematic uncertainty on the current best-fit oscillation parameters at T2K~\cite{OA2023}. Improved constraints, particularly on $\nu_e$ cross sections, are necessary to determine oscillation parameters to better precision at T2K, as well as at next-generation experiments including Hyper-Kamiokande~\cite{HK} and DUNE~\cite{DUNE}.

The $\nu_e \mathrm{CC} \pi^+$ channel has been observed in several inclusive $\nu_e$ cross-section measurements~\cite{Gargamelle-nueCC,T2K-nueCC,nueCC-antinueCC,NOVA-double-differential,coherent-nueCC}, as well as in an exclusive measurement on argon~\cite{MicroBooNE_nueCC1Pi+}. Simulations with the \textsc{Neut} event generator~\cite{NEUT} suggest single-pion production ($\mathrm{CC}1\pi^+$) channels contribute ${\sim}10\%$ to the $\nu_e$ appearance signal from the T2K flux given the current best-fit oscillation parameters \cite{OA2023}. Measuring pion production and $\nu_e$ cross sections at T2K presents unique challenges due to low statistics from the beam composition and numerous backgrounds, limiting the impact of these samples on constraining oscillation parameters. A recent measurement of $\nu_e\mathrm{CC}1\pi^+$, which combined T2K data with Super-Kamiokande atmospheric data, revealed an event rate excess localized to low lepton momentum~\cite{T2KSK_joint}. This study measured 225 events compared to a best-fit Monte Carlo (MC) prediction of $160 \pm 19$ under the near detector constraint. Similar excesses have also been reported using only T2K data, though with lower statistical significance \cite{OA2021,OA2023}. This analysis directly leverages data from the T2K off-axis near detector, ND280, to constrain the $\nu_e \mathrm{CC} \pi^+$ channel on carbon, improving our understanding of this crucial interaction.

The signal definition for this measurement is any CC interaction that produces an electron and at least one positively charged pion that escapes the nucleus. This is $\nu_{e}A \rightarrow e^- \pi^+ X$ where $A$ is a target nucleus and $X$ represents an arbitrary combination of hadrons, which, unlike the far detector samples, may include additional $\pi^+$. Phase space constraints are applied to the signal definition, ensuring the $e^-$ and $\pi^+$ can plausibly be detected and reliably reconstructed. The constraints are imposed on the $e^-$ momentum $\left(p_e \right)$ and angle with respect to the average neutrino direction $\left( \theta_e \right)$, as well as the pion momentum $\left(p_\pi \right)$; these are $0.35 < p_{e} < 30\;\text{GeV}/c$, $\cos{\theta_{e}} > 0.7$, and $p_{\pi} < 1.5\;\text{GeV}/c$.

\section{T2K experiment}
\label{sec:T2K}

Tōkai to Kamioka (T2K) is a long-baseline neutrino oscillation experiment in Japan~\cite{T2K}. The J-PARC facility houses both the neutrino beam-line and the near detector complex, which is situated 280\;m downstream from the neutrino beamline target. Super-Kamiokande (SK), a 50\;kt water Cherenkov detector located 295\;km to the west of J-PARC, serves as the far detector~\cite{Super-K}. A neutrino beam with peak energy 0.6\;GeV is produced by steering a beam of 30\;GeV protons towards a graphite target~\cite{flux}. This produces charged pions and kaons, which are then focused using magnetic horns towards a decay volume. Forward horn current (FHC) and reverse horn current (RHC) modes are used to select $\nu_\mu$ and $\bar{\nu}_\mu$ from decays of focused $\pi^\pm$ respectively. In FHC mode, the beam possesses an initial fractional composition of 92.6\% $\nu_\mu$ with 6.2\% $\bar{\nu}_\mu$, 1.1\% $\nu_e$ and 0.1\% $\bar{\nu}_e$~\cite{numuCC1Pi-H2O}.

ND280 is a tracking detector used to measure the unoscillated neutrino beam at the same off-axis angle as SK $\left( 2.5^{\circ} \right)$. The main target masses of ND280 are two fine-grained detectors (FGDs)~\cite{FGD}. These are comprised of modules made from polystyrene scintillator bars; each adjacent module has 192 bars oriented in alternating directions perpendicular to the neutrino beam. The upstream FGD (FGD1) has 15 modules and is predominantly carbon (86.1\%) by mass, with smaller proportions of hydrogen (7.4\%), oxygen (3.7\%), titanium (1.7\%), silicon (1.0\%), and nitrogen (0.1\%). Three time projection chambers (TPCs) adjacent to the FGDs are used to perform particle identification (PID) as well as for measuring the charge and momentum of traversing particles~\cite{TPC}. The TPCs and FGDs are surrounded by electromagnetic calorimeters (ECals)~\cite{ECal} which are used to detect neutral particles, perform track-shower separation and measure particle energy. The detector modules are enclosed by the 0.2\;T UA1/NOMAD magnet, instrumented with the scintillator-based Side Muon Range Detector~\cite{SMRD}. 

This analysis uses an exposure of $11.6 \times 10^{20}$ protons on target (POT) collected by ND280 between 2010-2013 and 2016-2017 in FHC mode. Neutrino interactions are primarily simulated using the \textsc{Neut} 5.4.0~\cite{Neut5.4} event generator. The \textsc{Nuisance2} framework~\cite{Nuisance} is used to obtain additional predictions from the \textsc{Genie} 3.4.2 ("AR23" tune)~\cite{Genie3.0} and \textsc{NuWro} 21.9.2~\cite{NUWRO} generators.

\begin{figure*}[t]
\begin{center}
\makebox[\textwidth]{
\begin{minipage}[b]{0.33\textwidth}
\includegraphics[trim={0.65cm 0 1cm 0.4cm},clip,width=1.0\textwidth]{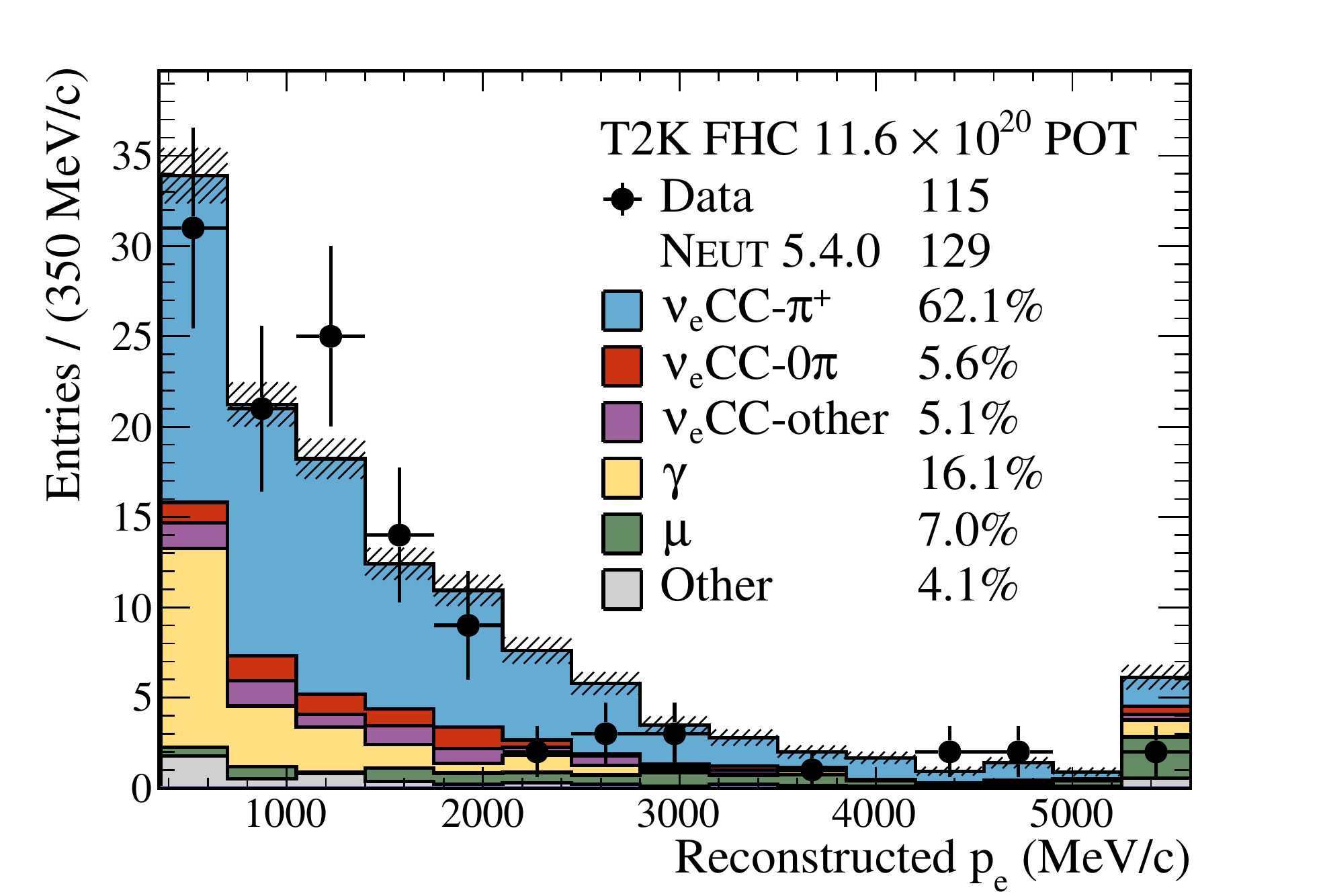}
\end{minipage}
\begin{minipage}[b]{0.33\textwidth}
\includegraphics[trim={0.65cm 0 1cm 0.4cm},clip,width=1.0\textwidth]{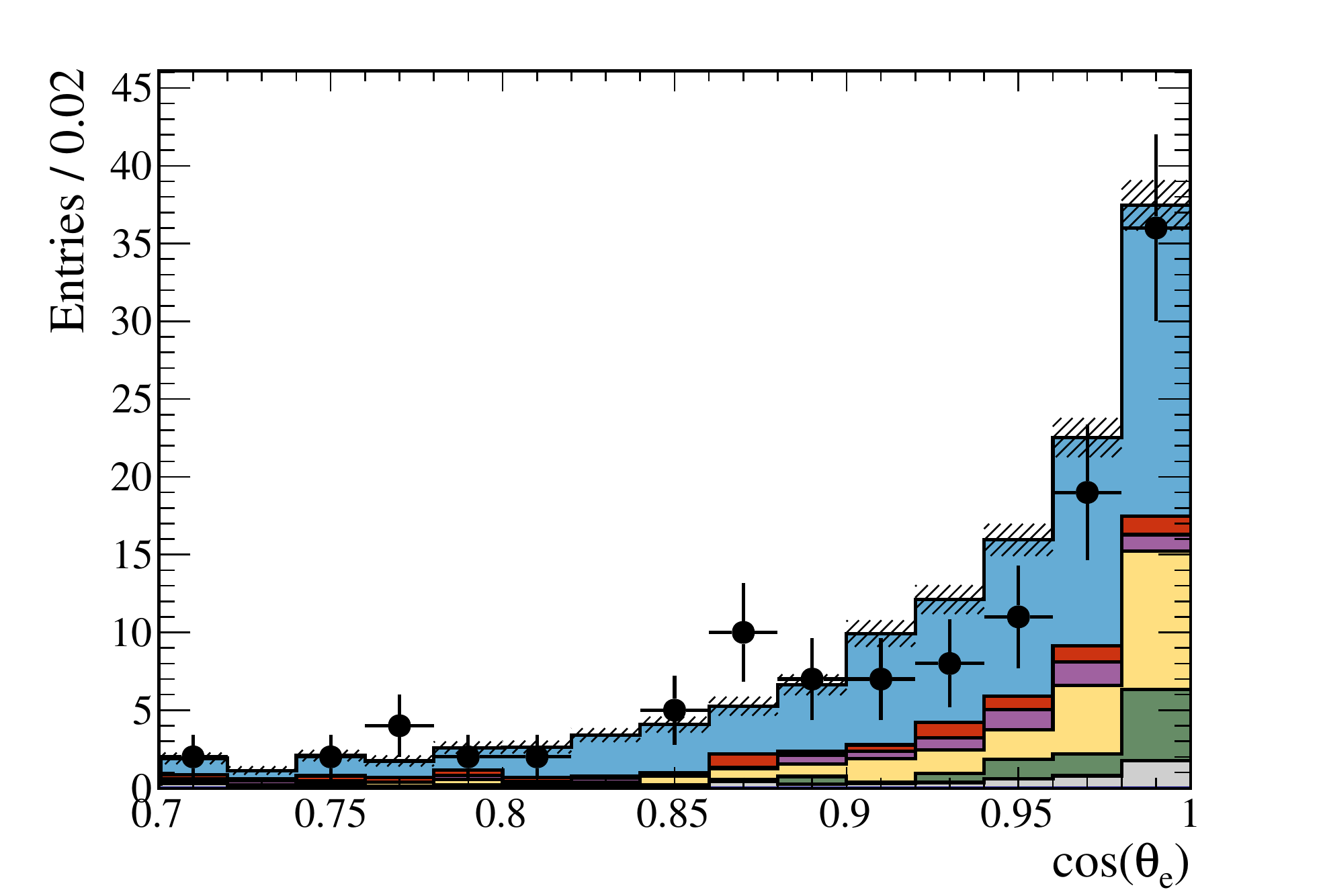}
\end{minipage}
\begin{minipage}[b]{0.33\textwidth}
\includegraphics[trim={0.65cm 0 1cm 0.4cm},clip,width=1.0\textwidth]{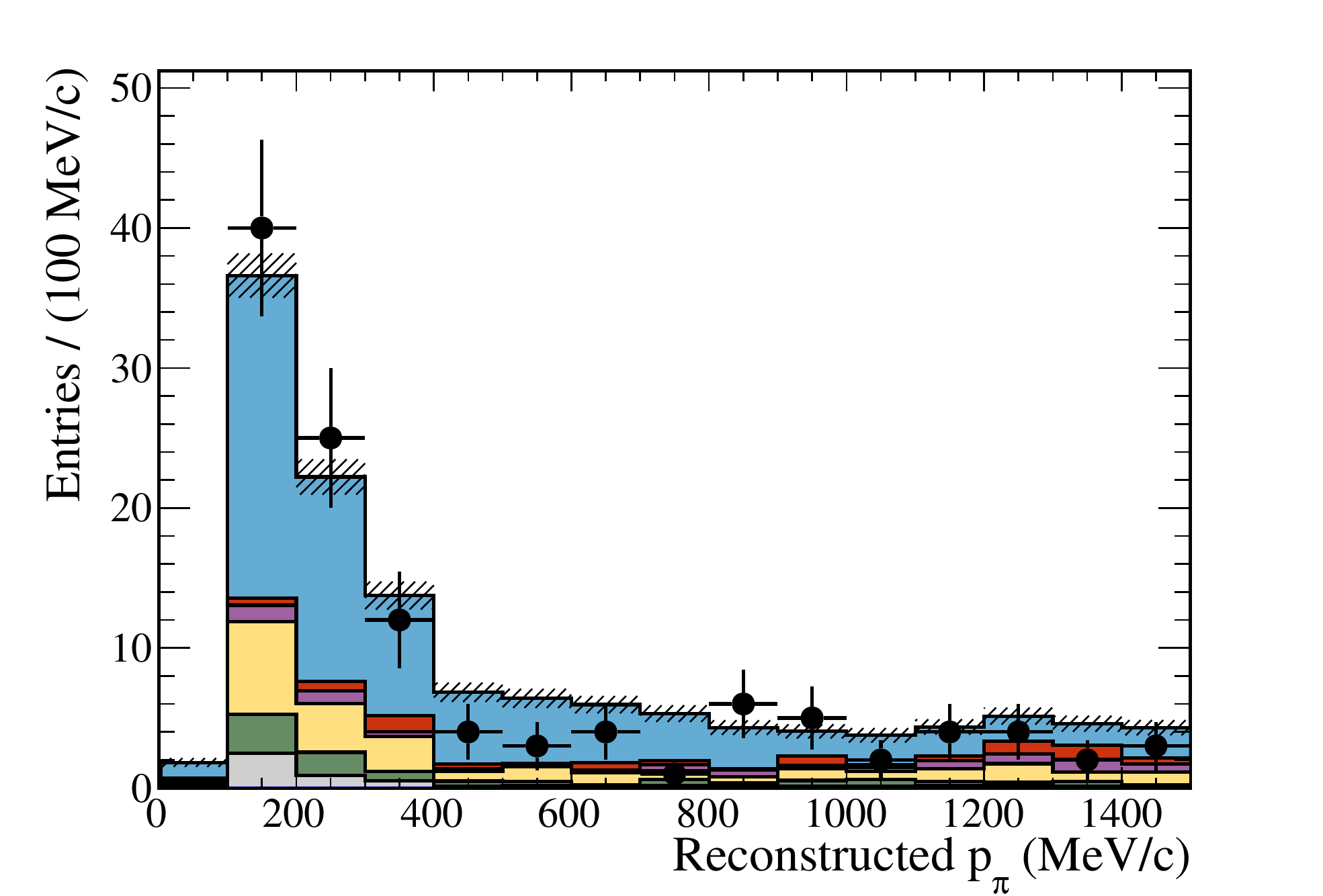}
\end{minipage}
}
\makebox[\textwidth]{
\begin{minipage}[b]{0.33\textwidth}
\includegraphics[trim={0.65cm 0 1cm 0.4cm},clip,width=1.0\textwidth]{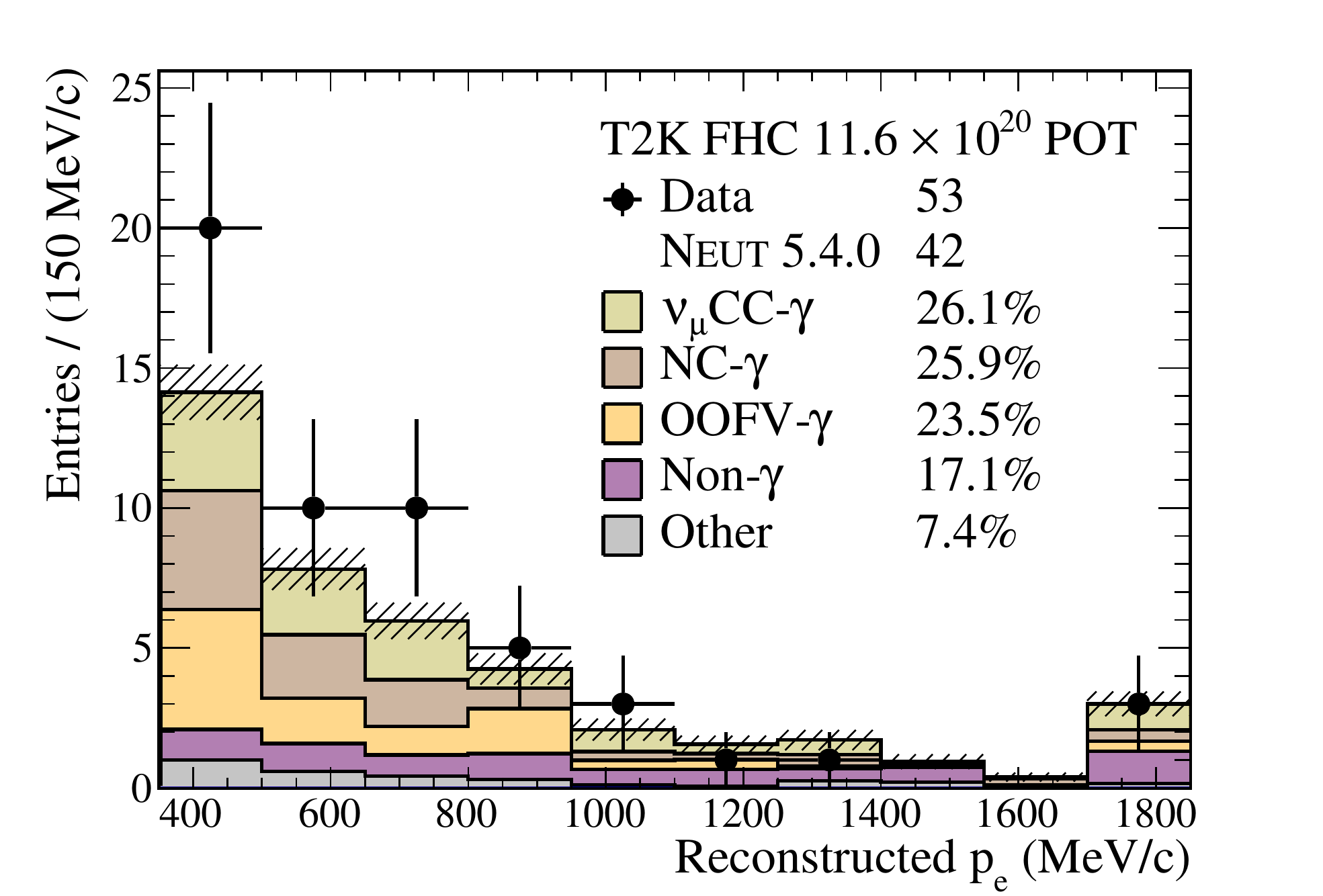}
\end{minipage}
\begin{minipage}[b]{0.33\textwidth}
\includegraphics[trim={0.65cm 0 1cm 0.4cm},clip,width=1.0\textwidth]{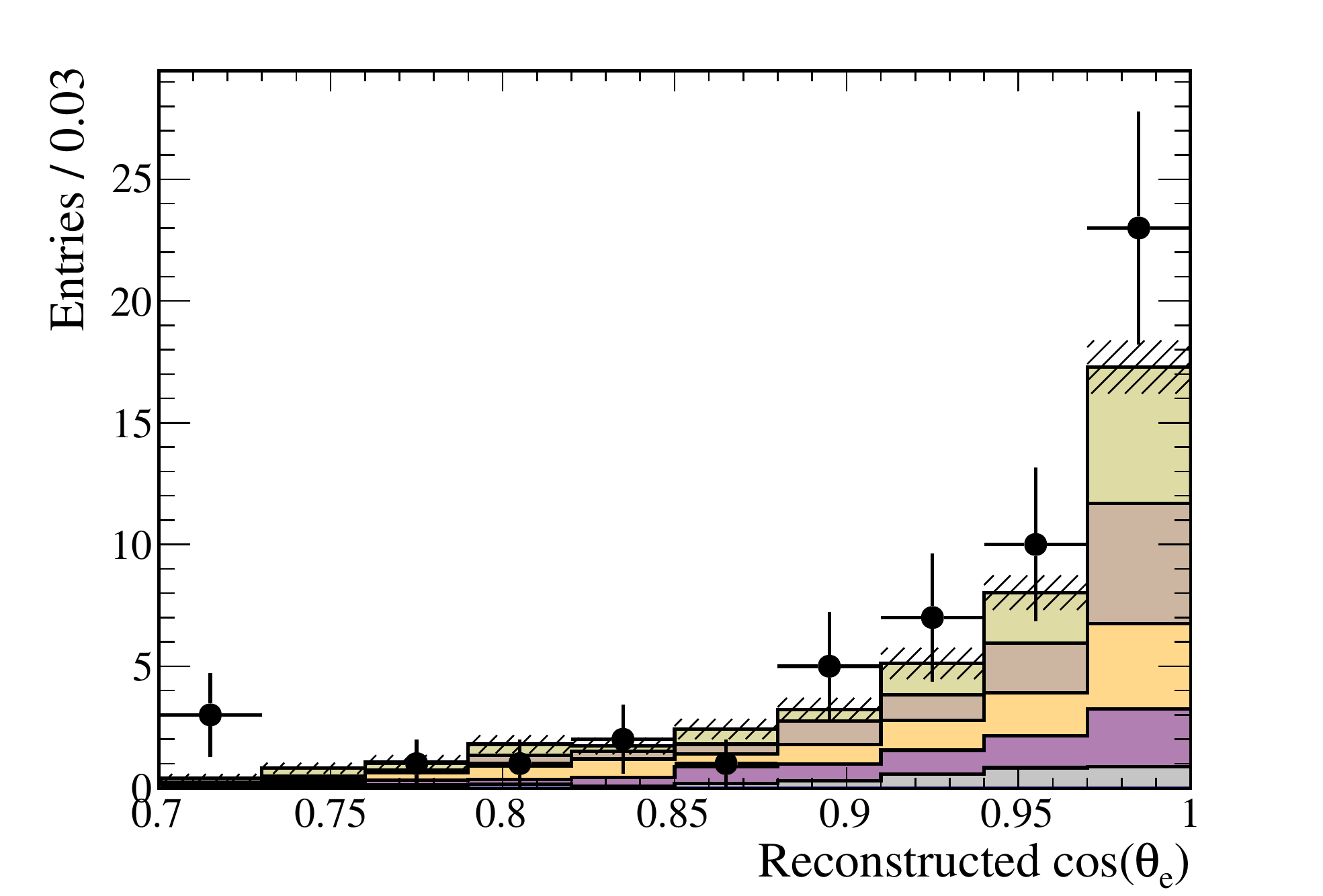}
\end{minipage}
\begin{minipage}[b]{0.33\textwidth}
\includegraphics[trim={0.65cm 0 1cm 0.4cm},clip,width=1.0\textwidth]{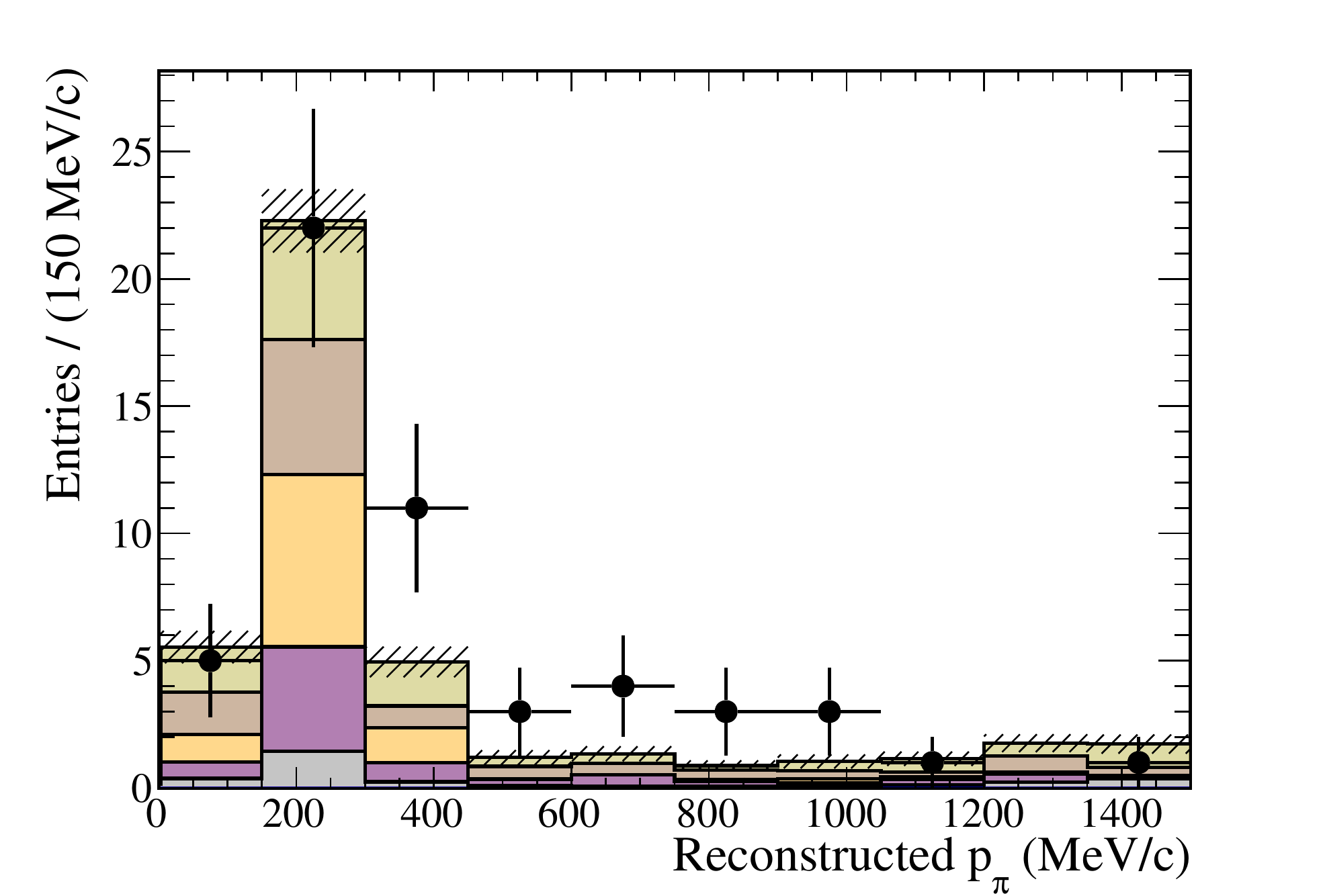}
\end{minipage}
}
\end{center}
\caption{\label{fig:RecoDistributions} Reconstructed electron and pion kinematics $(p_{e},\,\cos{\theta_{e}},\,p_{\pi})$ distributions for the signal samples (upper) and $\gamma$ controls samples (lower) in the restricted phase space. These show the number of selected data and pre-fit \textsc{Neut} MC events with statistical uncertainties, distinguished by true event type. \label{fig:observables}}
\end{figure*}

\section{Event selection}
\label{sec:Selection} 

Two signal-enriched samples with distinct pion detection methods are used to measure the $\nu_e \mathrm{CC}\pi^+$ event rate. The TPC-$\pi^+$ sample selects events where the pion passes from FGD1 into the downstream TPC (TPC2). The FGD-$\pi^+$ sample selects FGD1-contained pions that decay to Michel electrons by $\pi^+ \rightarrow \mu^+ \nu_\mu$ and $\mu^+ \rightarrow e^+ \bar{\nu}_{\mu} \nu_e$. The primary $\nu_e$ vertex is restricted to a fiducial volume (FV) within FGD1, excluding the five outermost scintillator bars in each layer where events tend to be mis-reconstructed, leaving a total target mass of 919.5\;kg~\cite{nueCC-antinueCC}.

The event selection developed for an inclusive $\nu_e$ cross-section measurement~\cite{nueCC-antinueCC} is used here, with the additional requirement that a pion is detected from the same vertex. In the TPC-$\pi^+$ sample, the leading positively charged track is subjected to PID cuts based on energy deposition $\left( dE/dx \right)$ in the TPC gas mixture, identifying $\pi^+$ candidates. The FGD-$\pi^+$ sample tags Michel electrons from clusters of delayed hits that tend to indicate $e^+$ production from $\pi^+$ decays. Both signal samples utilize sets of out-of-FV (OOFV) and ECal-track vetoes inherited from the inclusive analysis to reduce background contamination.

FGD-contained pion tracks are typically short, making their curvature difficult to reconstruct accurately, so their momenta cannot be determined using the same method as TPC tracks. The momentum of a FGD-contained $\pi^+$ is instead inferred from the location of hits left by Michel electrons with respect to the $\nu_e$ interaction vertex as described in Ref.~\cite{MEPion}. The relation between this distance, $d$ and the pion momentum is parameterized as $p_{\pi} = c_0 \times d^{c_1} + c_2$. The constants $c_{0,1,2}$ are determined from fitting the trend between $p_{\pi}$ and $d$ for MC. Recent ND280 cross section measurements have been limited to a phase space of $p_\pi \gtrsim 200\;\text{MeV}/c$ where pions can be reliably reconstructed in the TPCs~\cite{T2K-numuCC1Pi}; this technique removes this phase space limitation.

The largest background comes from $\pi^0$s that originate from the hadronic recoil system in $\nu_\mu$CC and $\nu_\mu$NC (neutral-current) interactions in FGD1; in the CC case the muons are below the tracking threshold. The $\pi^0$s decay into $\gamma\gamma$ pairs that convert to $e^+e^-$ pairs, initiating electromagnetic showers. In cases where the two $\gamma$s convert promptly and produce spatially separated track-like energy deposits above detection thresholds, they can be misidentified as a $\pi^+$ track and an electron shower emanating from a shared primary vertex. Around half of this background is removed by reconstructing the $e^+e^-$ invariant mass from the two candidate particles. The $e^+$ track is taken as the $\pi^+$ candidate from the TPC-$\pi^+$ sample, or as the next highest momentum positively charged track which passes electron PID for the FGD-$\pi^+$ sample. Events with an invariant mass clearly below the $e^+e^-$ peak ($m_{\text{inv}}<110\;\text{MeV}/c^2$) are removed. The $\gamma$ background remains dominant at low $p_e$ and high $\theta_e$, necessitating the aforementioned phase space constraints. The largest non-$\gamma$ background is from $\nu_e\mathrm{CC}$ events where a low momentum proton ejected from a nucleus is misidentified as a $\pi^+$; the remaining phase space constraint of $p_\pi > 1.5\;\text{GeV}/c$ reflects this inability to reliably identify tracks in this region. The next largest background is $\nu_{\mu}\mathrm{CC}$ events where the $\mu^-$ is above the tracking threshold and is misidentified as a $e^-$. The remaining backgrounds are small $\left( <~5\% \right)$ and are mostly those already discussed but with an OOFV primary vertex.

Events involving $\gamma$ conversions consistent with the $\gamma \rightarrow e^+e^-$ hypothesis $\left( m_{\text{inv}}~<~55.0\;\text{MeV}/c^{2} \right)$ are assigned to a pair of $\gamma$ control samples. These events must also pass one of the pion PID cuts used for the signal samples. The $\gamma$ control samples also use sets of OOFV vetoes, ensuring the normalized invariant hadronic mass and momentum transfer distributions of the selected $\pi^0$ production modes generally match those of the targeted backgrounds.

The data and MC events selected by the merged signal and control samples are shown projected as distributions of electron kinematics and pion momentum in Fig.~\ref{fig:RecoDistributions}. A total of 115 events are selected by the signal samples in the targeted phase space, compared with 129 events predicted by \textsc{Neut}. The signal samples have a combined efficiency of ${\sim}20\%$ that varies minimally across all in-phase space bins. The FGD-$\pi^+$ sample almost exclusively contributes to the lowest four $p_\pi$ bins. The data and MC shape distributions generally match well in most bins for the signal samples; the $\gamma$ control samples exhibit poorer agreement, but these are more statistically limited.

\begin{figure*}[t]
\begin{center}
\makebox[\textwidth]{
\begin{minipage}[b]{0.33\textwidth}
\includegraphics[trim={0.325cm 0cm 2.3cm 1.2cm},clip,width=1.0\textwidth]{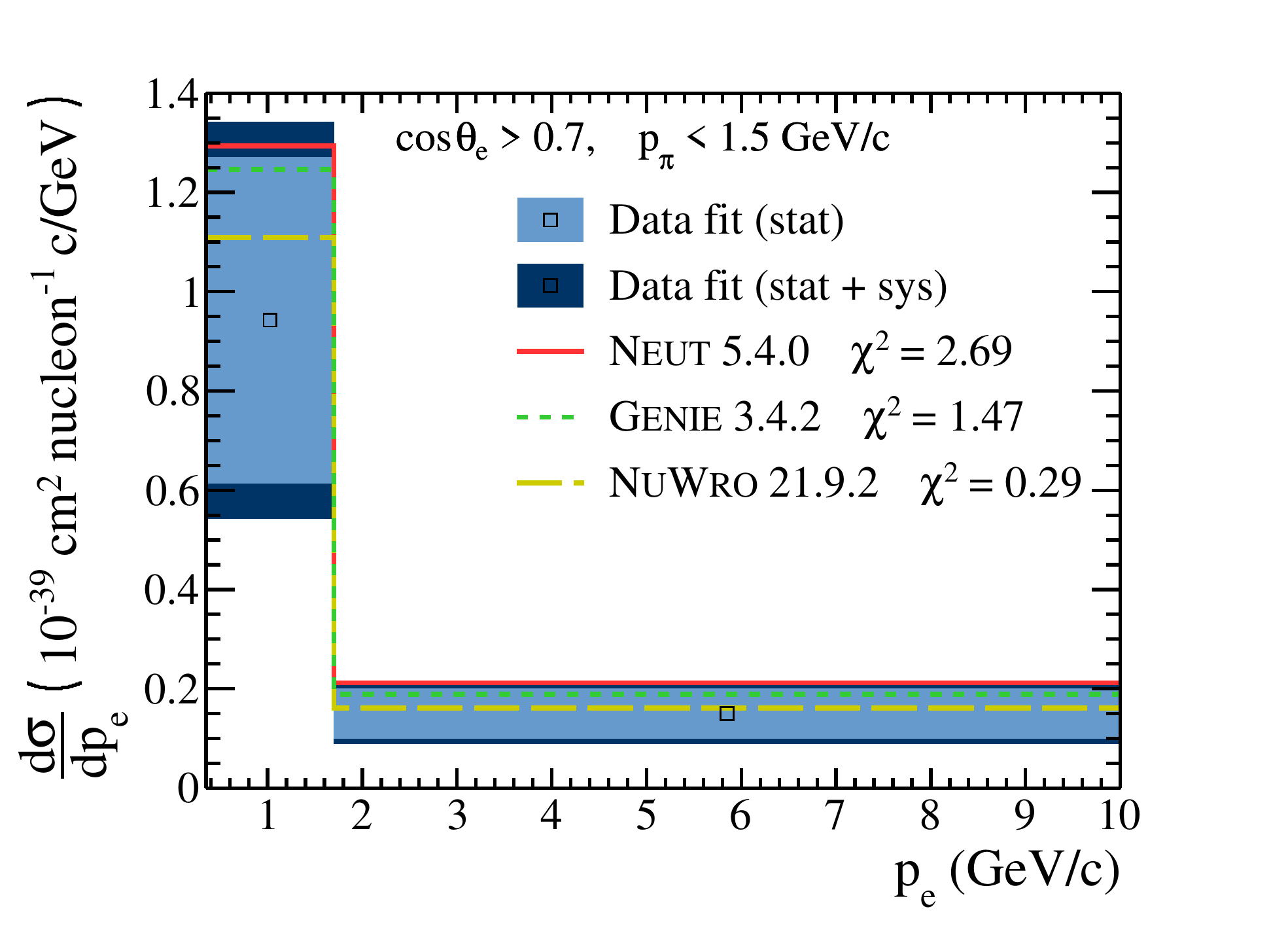}
\end{minipage}
\begin{minipage}[b]{0.33\textwidth}
\includegraphics[trim={0.325cm 0cm 2.3cm 1.2cm},clip,width=1.0\textwidth]{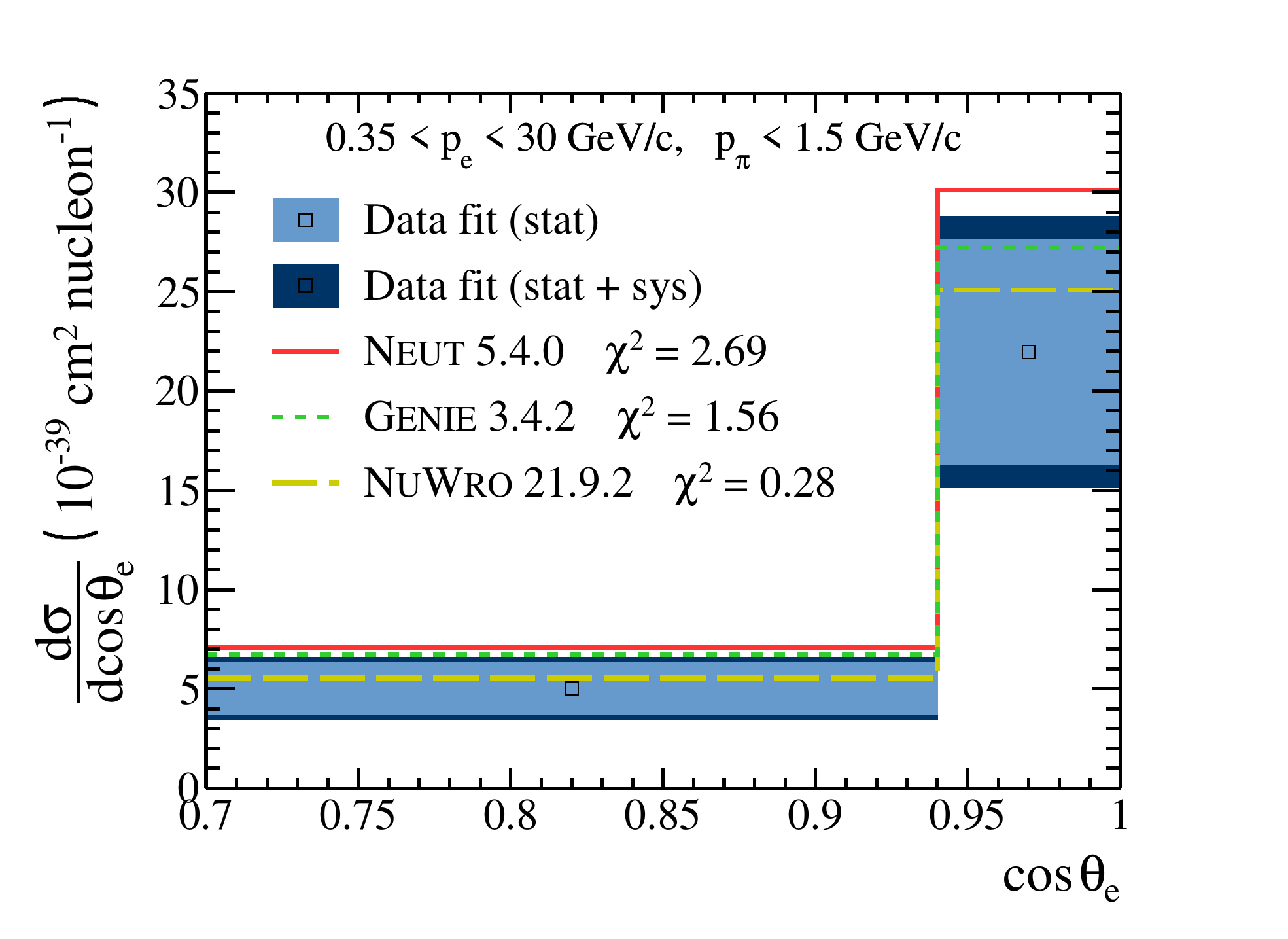}
\end{minipage}
\begin{minipage}[b]{0.33\textwidth}
\includegraphics[trim={0.325cm 0cm 2.3cm 1.2cm},clip,width=1.0\textwidth]{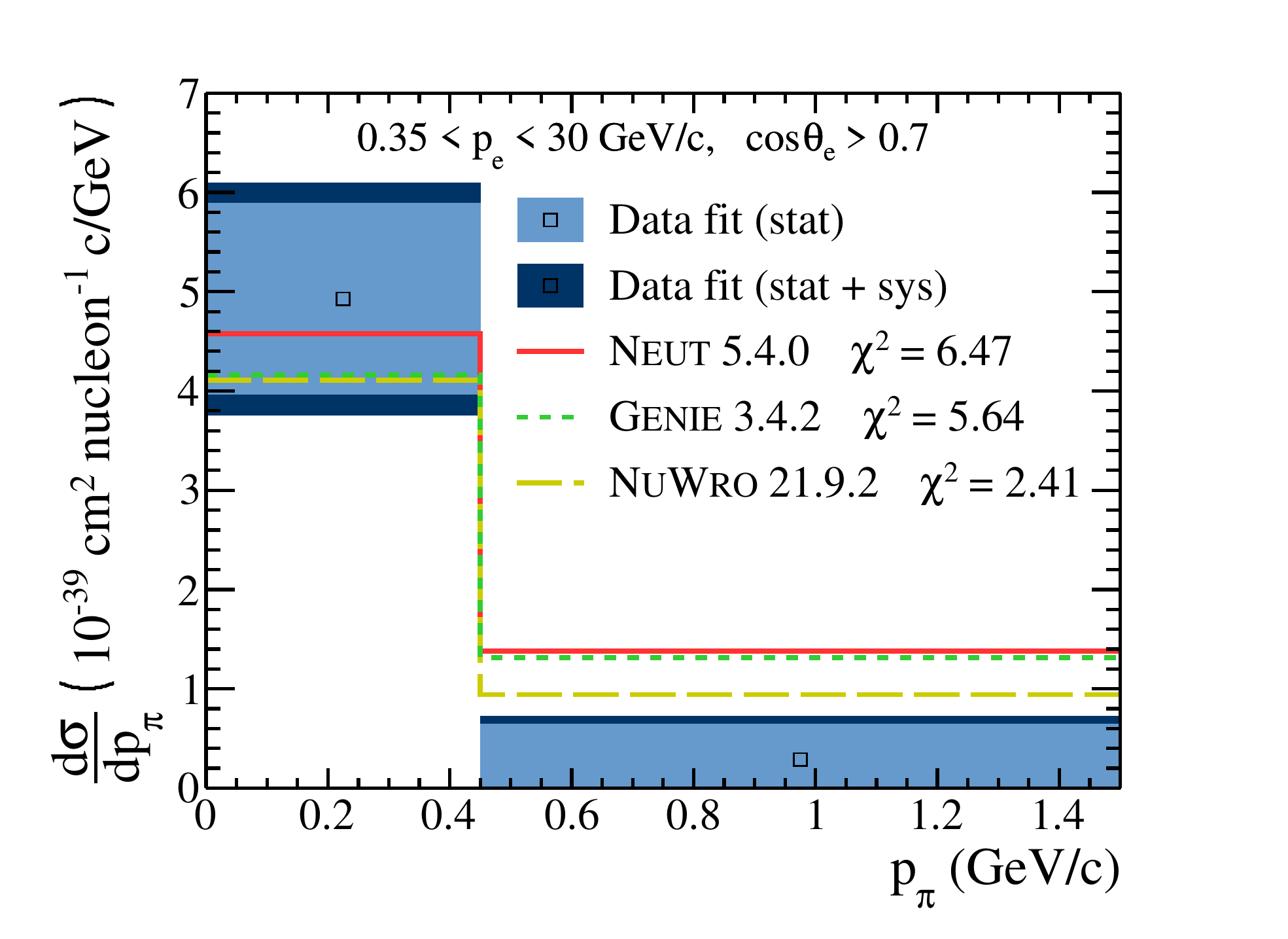}
\end{minipage}}
\end{center}
\caption{\label{fig:Results} The differential flux-integrated cross-section $\left(d\sigma\right)$ results and predictions from \textsc{Neut}, \textsc{Genie} and \textsc{NuWro} as a function of particle kinematics $(p_e,\,\cos{\theta_e},\,p_\pi)$. The upper $p_e$ bin extends to $30\;\text{GeV}/c$ and is normalized to account for the lower effective bin width. \label{fig:results}}
\end{figure*}

\section{Fit and extraction}
\label{sec:Fitting}

The unfolding method involves an unregularized binned template parameter likelihood fit used in several recent T2K cross section analyses~\cite{TKI, joint-fit, CC-COH}. The binning scheme uses eight signal-region bins that enclose the target phase space. The intermediate bin edges are optimized to approximately distribute an equal amount of signal events between the bins along each kinematic observable. The fit uses template parameters that are proportional to the signal event rate in each true kinematic bin; nuisance parameters are assigned to scale combinations of MC events corresponding to different systematic uncertainties. The template and nuisance parameters are iteratively varied to minimize the data-MC difference, yielding a set of best-fit parameters. The four samples are fitted separately and the systematic uncertainties are supplied as covariance matrices that encode the initial uncertainties of and correlations between nuisance parameters. The template parameters have no initial uncertainty and are free to move without prior constraints. During the fitting process, the negative log-likelihood ratio $\chi^{2} \simeq -2 \log{\left( \mathcal{L_{\text{stat}}} \right)} -2 \log{\left( \mathcal{L_{\text{sys}}} \right)}$ is minimized. The $\mathcal{L_{\text{stat}}}$ term is the likelihood ratio for the Poisson distribution with Barlow-Beeston modifications to account for finite MC statistics~\cite{Barlow}, and $\mathcal{L_{\text{sys}}}$ is the likelihood ratio for the set of parameter values at a specific fit point given their prior uncertainties divided by the maximum likelihood for those parameter values.

The differential cross section values are calculated with respect to a kinematic variable $x$ using $\mathrm{d}\sigma/\mathrm{d}x_i = \hat{N}_{i}/\epsilon_i \Phi N_T \Delta x_i$, where $\Delta x$ is an interval in 3D of the kinematic space defined by the analysis bin edges, $\hat{N}$ is the background subtracted number of signal events, $\epsilon$ is the selection efficiency, and $i$ denotes the bin index. This is normalized by the flux integral $\Phi$ and the number of nuclear targets $N_T$. The fitting process yields best-fit values and post-fit uncertainties for $\hat{N}_{i}$, $\epsilon_i$ and $\Phi$. The total cross section $\left( \sigma \right)$ is calculated by summing the differential results multiplied by their bin widths. The uncertainties on the binned cross sections are evaluated by varying the fit parameters within their post-fit uncertainties and numerically propagating these changes to the cross section results.

\section{Sources of systematic uncertainty}
\label{sec:Systematics}

The systematic uncertainties arise from the flux model, the detector response and the $\nu\text{-}A$ interaction model. The uncertainties on the $\nu_\mu$, $\bar{\nu}_\mu$, $\nu_e$ and $\bar{\nu}_e$ flux modes are encoded in a covariance matrix that constrains the total shape and normalization uncertainties of the flux in the neutrino energy-flavor space~\cite{flux}. The flux prediction is informed by replica target hardon production measurements by NA61/SHINE~\cite{NA61-1-2, NA61-2-2}. The uncertainties in the detector response are determined using separate dedicated control samples each corresponding to different aspects of event reconstruction in ND280. The difference in event rates between data and MC are then used to evaluate each source of uncertainty; these are also encapsulated by a single covariance matrix containing the reconstructed space bins of the four analysis samples. The dominant uncertainties in the detector response for the signal samples are from the TPC PID and matching between TPC-ECal tracks. For the $\gamma$ control samples, the detector mass uncertainty effect on the photon mean free path is dominant. The $\nu\text{-}A$ interaction model uncertainties are theory-driven and correspond to parameters that model signal and background interactions as well as final-state interactions (FSI). A  description of each parameter and the pre-fit uncertainties assumed is available in Ref.~\cite{OA2023}. For this analysis, the dominant interaction model uncertainties are those associated with FSI and normalizations to the multi-pion production channels. The fractional systematic and statistical uncertainties on the total cross section are shown in Table~\ref{table:Uncertainties}.

\begin{table}[h!tb]
\renewcommand{\arraystretch}{1.2}
\begin{tabularx}{0.9\columnwidth}{>{\raggedright\arraybackslash}X>{\centering\arraybackslash}X}
\hline
\hline
Uncertainty source & Fractional error on $\sigma$ [\%] \\
\hline
Detector response & 5.7 \\
Flux model & 6.8 \\
Interaction model & 7.8 \\
Target mass & 0.7 \\
\hline
Total statistical & 20.6 \\
Total systematic & 11.7 \\
\hline
\hline
\end{tabularx}
\newline \newline
\caption{The contributions of each source of uncertainty on the total flux-integrated cross section $\left( \sigma \right)$. \label{table:Uncertainties}}
\end{table}

\section{Results}
\label{sec:Results}
The total flux-integrated cross section is $[2.52 \pm 0.52\;(\text{stat}) \pm 0.30\;(\text{sys})] \times 10^{-39}\;\text{cm}^{2}\;\text{nucleon}^{-1}$. The predicted total flux-integrated cross sections from event generators are listed alongside this in Table~\ref{table:Results}. All of the neutrino event generators predict a larger cross section than this measurement, an effect which ranges from 0.5--1.6$\sigma$. The differential cross section results are shown in Fig.~\ref{fig:Results}; these results are also lower than the event generator predictions in the strongly forward going and high pion momentum regions of kinematic phase space. The cross section result in the $0.45 < p_\pi < 1.5\;\text{GeV}/c$ region exhibits the largest discrepancy, with the \textsc{Neut} and \textsc{Genie} predictions overestimating this by $2.5\sigma$ and $2.4\sigma$ respectively. The \textsc{NuWro} prediction is notably closer but still $>\!1\sigma$ from the measured cross section. These results contrast with a $\nu_\mu \mathrm{CC}1\pi^+$ measurement at T2K \cite{numuCC1Pi-H2O}, where the data is overestimated by \textsc{Neut} and \textsc{Genie} below $0.2\;\text{GeV}/c$, while the region above this exhibits good agreement in most bins. From Fig.~\ref{fig:RecoDistributions}, this ND280 sample does not experience an event rate excess comparable to what is seen in the far detector analyses~\cite{OA2021,T2KSK_joint}. However, this analysis does not measure much of the phase space relevant to these samples $(p_e < 0.35\;\text{GeV}/c)$. Below this limit it is too difficult to distinguish between signal events and the $\gamma$ backgrounds; ascertaining whether any discrepancies in this region are a result of mismodelling the signal or background is not possible.

\begin{table}[h!tb]
\renewcommand{\arraystretch}{1.2}
\begin{tabularx}{0.9\columnwidth}{>{\raggedright\arraybackslash}X>{\raggedright\arraybackslash}X>{\centering\arraybackslash}X}
\hline \hline
Generator & $\sigma\;(10^{-39}\;\text{cm}^{2}\;\text{nucl}^{-1})$ & $p$-value \\ 
\hline
\textsc{Neut} 5.4.0 & 3.51 & 0.30 \\ 
\textsc{Genie} 3.4.2 & 3.25 & 0.59 \\ 
\textsc{NuWro} 21.9.2 & 2.84 & 0.89 \\ 
\hline
Data & $2.52 \pm 0.60$ & - \\ 
\hline \hline
\end{tabularx}
\newline \newline
\caption{The measured and predicted total flux-integrated cross sections $\left( \sigma \right)$ per target nucleon (nucl) from \textsc{Neut}, \textsc{Genie} and \textsc{NuWro}. The $p$-value is calculated using the total $\chi^{2}$ between the three-dimensionally binned data and Monte Carlo histograms for all eight in-phase space bins. \label{table:Results}}
\end{table}

\section{Conclusions}
\label{sec:Conclusions}

The T2K collaboration has performed the first cross section measurement of $\nu_e \mathrm{CC} \pi^+$ on carbon in a restricted kinematic phase space. The total flux-integrated result is lower than predictions of the \textsc{Neut}, \textsc{Genie} and \textsc{NuWro} MC event generators. The cross section result for $0.45 < p_\pi < 1.5\;\text{GeV}/c$ exhibits a more substantial disagreement with \textsc{Neut} and \textsc{Genie}. This result represents an important first step in testing $\nu_{e}\text{-}A$ interaction models for the exclusive $\pi^+$ production channel. Future studies should focus on enhancing $e^\pm$-$\gamma$ separation techniques to reduce the $\pi^0$ backgrounds in the low $p_e$ region. The recent upgrades to ND280 may be capable of addressing this more effectively by making full use of its highly-segmented FGD \cite{NDUP, SFGD}. Larger datasets will also be necessary to reduce the statistical uncertainty and more precisely evaluate the performance of event generators in describing this interaction.

\titleformat{\section}[block]{\scshape\bfseries\centering}{}{}{}[]
\titlespacing*{\section}{0pt}{5ex}{2ex}

\section{Acknowledgements}
\label{sec:Acknowledgements}

\pretolerance=9000

\indent The T2K collaboration would like to thank the J-PARC staff for superb accelerator performance. We thank the CERN NA61/SHINE Collaboration for providing valuable particle production data. We acknowledge the support of MEXT,   JSPS KAKENHI (JP16H06288, JP18K03682, JP18H03701, JP18H05537, JP19J01119, JP19J22440, JP19J22258, JP20H00162, JP20H00149, JP20J20304) and bilateral programs (JPJSBP120204806, JPJSBP120209601),  Japan; NSERC, the NRC, and CFI, Canada; the CEA and CNRS/IN2P3, France; the Deutsche Forschungsgemeinschaft (DFG, German Research Foundation) 397763730, 517206441, Germany; the NKFIH (NKFIH 137812 and TKP2021-NKTA-64), Hungary; the INFN, Italy; the Ministry of Science and Higher Education (2023/WK/04) and the National Science Centre (UMO-2018/30/E/ST2/00441, UMO-2022/46/E/ST2/00336 and UMO-2021/43/D/ST2/01504), Poland;  the RSF (RSF 24-12-00271) and the Ministry of Science and Higher Education, Russia; MICINN  (PID2022-136297NB-I00 /AEI/10.13039/501100011033/ FEDER, UE, PID2021-124050NB-C31, PID2020-114687GB-I00,  PID2019-104676GB-C33), Government of Andalucia (FQM160, SOMM17/6105/UGR) and the University of Tokyo ICRR's Inter-University Research Program FY2023 Ref. J1, and ERDF and European Union NextGenerationEU funds (PRTR-C17.I1) and CERCA program, and University of Seville grant Ref. VIIPPIT-2023-V.4, and Secretariat for Universities and Research of the Ministry of Business and Knowledge of the Government of Catalonia and the European Social Fund (2022FI\_B 00336), Spain; the SNSF and SERI (200021\_185012, 200020\_188533, 20FL21\_186178I), Switzerland; the STFC and UKRI, UK; the DOE, USA; and NAFOSTED (103.99-2023.144,IZVSZ2.203433), Vietnam. We also thank CERN for the UA1/NOMAD magnet, DESY for the HERA-B magnet mover system, the BC DRI Group, Prairie DRI Group, ACENET, SciNet, and CalculQuebec consortia in the Digital Research Alliance of Canada, and GridPP in the United Kingdom, and the CNRS/IN2P3 Computing Center in France and NERSC (HEP-ERCAP0028625). In addition, the participation of individual researchers and institutions has been further supported by funds from the ERC (FP7), “la Caixa” Foundation  (ID 100010434, fellowship code LCF/BQ/IN17/11620050), the European Union’s Horizon 2020 Research and Innovation Programme under the Marie Sklodowska-Curie grant agreement numbers 713673 and 754496, and H2020 grant numbers  RISE-GA822070-JENNIFER2 2020 and RISE-GA872549-SK2HK; the JSPS, Japan; the Royal Society, UK; French ANR grant number ANR-19-CE31-0001 and ANR-21-CE31-0008; and  Sorbonne Université Emergences programmes; the SNF Eccellenza grant number PCEFP2\_203261;  the VAST-JSPS (No. QTJP01.02/20-22);  and the DOE Early Career programme, USA. For the purposes of open access, the authors have applied a Creative Commons Attribution license to any Author Accepted Manuscript version arising.

\bibliography{bibliography}

\end{document}